\documentclass[aps,twocolumn,prl]{revtex4}
\usepackage{color}

\usepackage{epsfig}
\usepackage{epstopdf}
\usepackage{amsmath}
\usepackage[latin1]{inputenc}
\usepackage{graphicx}

\newcommand{\nc}{\newcommand}
 \nc{\tcb}{\textcolor{blue}}  
 \nc{\tcr}{\textcolor{red}}
 \nc{\be}{\begin{equation}} 
 \nc{\ee}{\end{equation}}
 \nc{\bea}{\begin{eqnarray}}  
 \nc{\eea}{\end{eqnarray}}
 \nc{\ba}{\begin{array}}  
 \nc{\ea}{\end{array}}
 \nc{\rds}{{\rm d}s} 
 \nc{\rdt}{{\rm d}t} 
 \nc{\rdr}{{\rm d}r}
 \nc{\rdO}{{\rm d}\Omega} 
 \nc{\s}{{\rm S}} 
 \nc{\Pl}{{\rm Planck}}
 \nc{\dis}{\displaystyle} 
 \nc{\crit}{_{\rm cr}} 
 \nc{\rd}{{\rm d}}
 \nc{\munu}{{\mu\nu}} 
 \nc{\erm}{{\rm e}}
 \nc{\drm}{{\rm d}}
 \nc{\ov}{\overline}
\newcommand{\sigmabf}{\mbox{\boldmath $\sigma$}}

\begin{document}

\title{Transition from Majorana to Weyl fermions and anapole moments}

\author{S. Esposito}
\email{sesposit@na.infn.it}%
\affiliation{
Istituto Nazionale di Fisica Nucleare, Sezione di Napoli, Complesso Universitario di Monte
S.\,Angelo, via Cinthia, I-80126 Naples, Italy}

\

\

\begin{abstract}
\noindent The apparent splitting of zero-bias conductance peaks, apparently observed in recent experiments concerning Majorana fermions in nanowires coupled to superconductors, can be interpreted as a manifestation of a transition in the structure from (massive) Majorana to (massless) Weyl fermions. A modification of the experiments in order to test such phenomenon is proposed by making recourse to the only possible electromagnetic interaction allowed to (massive) Majorana particles and mediated by their anapole moment. In suitably designed heterostructures with toroidal symmetry, the additional anapole interaction manifests itself in the lowering of the critical magnetic field required for the appearance of the zero-bias peak, a shift that can be directly measured and thus reveal the presence of Majorana (rather than Weyl) fermions. Anapole interactions, though not previously considered, may also be a powerful method to control the dynamics of Majorana fermions, and then to manipulate the qubit state in quantum computation.

\end{abstract}

\maketitle

\noindent A large effort has been devoted in recent times to the study of low-energy excitations in condensed matter systems described by a Dirac-like equation or its extensions or deformations, such as graphene \cite{diracgraphene}. Indeed, excitations in solids above the ground state may carry quantized amounts of energy, momentum, spin and electric charge, thus behaving like elementary particles: the positive and negative energy solutions of the Dirac equation, for example, model free conduction electrons and bound valence electrons, respectively. Even more exciting, given the possibility of implementing non-Abelian statistics relevant for topological quantum computing, is the realization of {\it Majorana fermions}, described by a particular Dirac equation whose solutions are real \cite{EM37}, so that they represent neutral, spin-1/2 particles that are their own antiparticles. In condensed matter physics, they arise as exotic superconducting states \cite{Majorana}; indeed, although electrons and holes have opposite charge $\pm e$, the charge difference $2 e$ can be absorbed as a Cooper pair in the dense superconducting condensate, so that at the Fermi level ($E=0$), in the middle of the superconducting gap, eigenstates are charge-neutral superpositions of electrons and holes. These states emerge in a variety of systems, ranging from $p$-wave superconductors, where electrons with the same spins pair up, 
to particular superfluids and quantum Hall states. 

More in general, since long time \cite{Rebbi,Rossi} it is known that fermion zero-modes are bound to topological defects (for example, certain vortex excitations in given systems support Majorana fermions residing inside their cores), and a number of proposals exists aimed at realizing Majorana zero-modes using heterostructures of semiconductor and superconductor, supeconductor and ferromagnet, quantum (anomalous) Hall state and superconductor, etc. \cite{SauLut,Alicea,FuKane}. Indeed, spin-triplet ($p_x+i p_y$) superconductivity supports zero-energy states appearing in half-quantum vortices, this being effectively equivalent to a full-quantum vortex in spinless ($p_x+i p_y$) superfluid or superconductor. However, Fu and Kane \cite{FuKane} also showed that when an ordinary ($s$-wave) superconductor is attached to a material with large spin-orbit coupling, due to the proximity of the two materials, Cooper pairs leak through the interface into the second material, thus inheriting superconductivity over (at least) a coherence length. Majorana fermions can thus be created in the vortices of $s$-wave superconductors deposited on the surface of a three-dimensional topological insulator, which exhibit a bulk gap with gapless surface states protected by topology, but are present as well in semiconductor nanowires in close proximity to $s$-wave superconductors, provided the semiconductor has a pronounced spin-orbit coupling and a magnetic field is applied along the wire in order to make the band appear spinless \cite{SauLut}.

Very recently, the scientific community got impressed by several claims about experimental signatures of Majorana fermions, observed just in some of such condensed systems \cite{Williams, Mourik, Rokhinson, Das, Deng}. For example, a magnetic field has been applied parallel to a one-dimensional (semiconducting) wire of InSb connected to a normal (not superconducting) metal gold electrode on one side of a circuit, whose other side was made of (superconducting) NbTiN \cite{Mourik}. One gate voltage applied to the semiconducting nanowire varied electron density, and thus created a barrier between normal and superconducting contacts, through which normal gold electrons could in case tunnel into the wire. However, a normal electron cannot usually tunnel into a superconducting gap, unless they tunnel into zero-energy Majorana states residing there. In such a case, electrons add to circuit's conductance, so that when a bias voltage between the normal metal and superconductor is applied for measuring the supercurrent conductance through the wire, a peak should develop for zero-bias voltage. This is just what has been apparently observed \cite{Mourik} (see also Ref. \cite{Das}).

While further experiments are required in order to confirm such results, here we focus on a often overlooked key point of the phenomenology of Majorana fermons, inspired by a peculiar observation that have come out from (some of) the mentioned experiments \cite{Lin}. 

As we have seen, a necessary condition for the manifestation of Majorana modes in the systems at hand, is the presence of a robust zero-bias peak in the differential tunneling conductance \cite{SauLut,Oreg,Flen,Seng,Law} in the middle of the superconducting gap. This occurrence is the direct consequence of the appearance of zero-energy modes localized at the two ends of a nanowire, which has become an effective chiral $p$-wave topological superconductor \cite{Seng,Kitaev}. Now, the required superconducting properties of the wire are produced by the proximity of the superconducting electrode, which induces a non-zero gap $\Delta$, while the topological phase has to be generated by a sufficiently large magnetic field $\bf B$ applied along the axis of the nanowire, provided that the associated Zeeman energy $E_Z=g \mu_B B/2$ satisfies the condition 
\be \label{TS}
E_Z > \sqrt{\Delta^2 + \mu^2} , 
\ee
where $\mu$ is the chemical potential of the wire \cite{SauLut,Oreg}. This is, then, the basic relation to be fulfilled; it is usual to refer directly to it in order to have Majorana zero-modes inside vortices located at the two ends of the nanowire. Indeed, in Ref. \cite{Mourik} \footnote{For definiteness, we refer only to the experimental observations reported in this paper; similar results, however, apply also to what reported in Ref. \cite{Das}.} the signal for the presence of Majorana modes was just the appearance of a zero-bias peak for a {\it non-zero} magnetic field ($B \, {\small \gtrsim} \, 0.1$ T), thus satisfying the condition (\ref{TS}). However, it has been pointed out that at higher values of the magnetic field ($B \, {\small \gtrsim} \, 0.5$ T), an apparent, unexpected splitting of the single zero-bias peak to a pair of narrow peaks (away from the Andreev bound state resonances) occurs. In Ref. \cite{Mourik}, this has been interpreted as the manifestation of a coupling of two nearby Majorana modes, as later confirmed by general calculations in Ref. \cite{Lin}. Since the length of the wire is finite, in fact, the possibility exists that the two Majorana modes localized at the two ends of the wire hybridize, as a consequence of an intravortex tunneling process \cite{Cheng}. The given explanation is, then, an acceptable interpretation of the experimental observations.

However, here we prefer to address the question from a more fundamental point of view. 

The most simple description of a spin-1/2 fermion is that of a massless particle obeying the Weyl equation, a special case of the Dirac equation with vanishing mass. The particle state can be associated to one definite spin (or, rather, helicity) state, while the antiparticle state to the other one. In terms of chiral spinors, which are the most convenient representation in the present case, they are described by two independent couples of state functions, which we label as follows ($C$ represents charge conjugation):
\be \label{left}
\psi_L , \quad \left( \psi_L \right)^C = \psi^C_{\, R}
\ee
and
\be \label{right}
\psi_R  , \quad \left( \psi_R \right)^C = \psi^C_{\, L}
\ee
These spinors are helicity eigenstates: $\psi_R$ [$\psi^C_{\, R}$] describes a right-handed particle [antiparticle] with positive helicity, while $\psi_L$ [$\psi^C_{\, L}$] describes a left-handed particle [antiparticle] with negative helicity. Since, for massless particles, the dynamical evolution of the first set 
is completely decoupled from that of the second one, 
\be \label{weyl}
\frac{\sigmabf \cdot {\mathbf p}}{E} \psi_R = \psi_R , \quad \frac{\sigmabf \cdot {\mathbf p}}{E} \psi_L = - \psi_L ,
\ee
($\sigmabf$ are Pauli matrices, while $\bf p$,$E$ denote the momentum, energy of the particle) it is possible to have situations where, for example, only the states in (\ref{left}) appear, or even have a non-equal mixture of states in (\ref{left}) and (\ref{right}). In such cases, parity is obviously violated. In condensed matter systems, this can be realized, for example, by a 3D Weyl semimetal phase in a topological insulator multilayer \cite{Burkov} or in the Sr$_2$RuO$_4$ chiral superconductor \cite{Nobukane}.

For massive particles, the mentioned decoupling no longer holds, since the Dirac equation mixes right-handed and left-handed states:
\be \label{dirac}
\left( E - \sigmabf \cdot {\mathbf p} \right) \psi_R = m \psi_L , \quad \left( E + \sigmabf \cdot {\mathbf p} \right) \psi_L = m \psi_R .
\ee
Thus, we have now only one independent quantity describing the particle, composed by four components, and it is common to introduce the Dirac spinor as a linear combination of chiral spinors, $\psi = \psi_L + \psi_R$ (and similarly for $\psi^C$), in terms of which the Dirac equation is written in the more usual form \footnote{Of course, the following equation, as the previous ones, is for a ``free particle"; if other energy sources are present, they sum to the diagonal terms in Eq. (\ref{diracm}).}: 
\be \label{diracm}
\left[ \ba{cc} \sigmabf \cdot {\mathbf p}  & m \\ m & - \sigmabf \cdot {\mathbf p} \ea \right] \psi = E \psi .
\ee
Although chiral spinors violate parity, their combinations in the Dirac spinors evidently do not. In solid state systems, the mass term is mimicked by either a constant value, which produce a gap in the energy spectrum, or a inhomogenous term that varies with position, as for example the phonon field $\phi({\bf r})$; in the latter case, a topologically non-trivial configuration, such as that of a vortex, would support a zero-energy mode. Dirac-like excitations are present in a variety of systems, among which the most famous one is graphene \cite{diracgraphene}, displaying also a number of unusual effects, such as the emergence of a fractional charge. Against a certain confusion existing in the condensed matter literature, it should be stressed that massless Dirac fermions are completely {\it equivalent} to Weyl fermions, as coming out particularly evident from the above: a four-component state function splits into two independent two-component state functions. However, in solid state systems, this evenience does not lead directly to parity violation, as already discussed, depending on the physical realization of the fermionic excitations in the given material.

However, the mentioned description of massive fermions is not the only possible one. Indeed, the same equation (\ref{diracm}) may be built if the Dirac spinor $\psi$, made up of $\psi_L$ and $\psi_R$ chiral states, is replaced by the spinor $\psi_{M1} = \psi_L + \psi^C_{\, R}$ or $\psi_{M2} = \psi_L - \psi^C_{\, R}$, made up of {\it only} the chiral state $\psi_L$ and its charge conjugated one (analogous terms may be built with the other chiral state $\psi_R$). In such cases, it follows immediately that $\left( \psi_M \right)^C = \pm \, \psi_M$, that is we have self-charge conjugated states describing Majorana fermions. Although, as in the Dirac case, we have again formally a four-component state function, in a sense (as often appeared in the literature) a Majorana fermion is ``half" a Dirac fermion, since it is composed by just half the degrees of freedom of a Dirac particle. The situation is, however, completely different from that of Weyl fermions, since the specific properties are different: Majorana fermions are massive, chargeless and parity is not at all violated. As above, in the limit of zero mass, Majorana fermions reduce, obviously, to Weyl fermions. As already mentioned, when the mass term is replaced by an inhomogeneous vortex profile, for example, zero-energy solutions are possible \cite{Rossi}, as in the system formed by a superconductor in contact with a topological insulator \cite{FuKane}.

Let us now come back to the problem of the splitting of the zero-bias peak in the experiments discussed above, which has benn attributed to the superposition of the two Majorana modes at the ends of the nanowire, and assume for simplicity (for the moment) that such superposition is maximal. According to the notation introduced, the two Majorana fermions in the nanowire are described \cite{Kitaev} by the spinors denoted with $\psi_{M1}$ and $\psi_{M2}$, so that the excitations resulting from the maximal superposition of the two modes are described by $\psi_{M1} + \psi_{M2}$ or $\psi_{M1} - \psi_{M2}$. However, it immediately follows that such states are simply chiral Weyl states: in our scheme, then, the splitting of the zero-bias peak is predicted to be just the result of a transition from Majorana to Weyl modes. 

Is this generally compatible with observations? First of all, we note that a splitting has been observed into only {\it two} peaks, and this could be a manifestation of the fact that the degrees of freedom of a Weyl fermion are just half those of a Majorana fermion. Moreover, a transition from Majorana to Weyl fermions means, as recalled above, the vanishing of the mass term in (\ref{diracm}) which, in the system considered, translates into the effective superconducting gap experienced by Majoranas in the nanowire. Now, the splitting mentioned has been observed for higher values of the magnetic field along the wire, the condition (\ref{TS}) being still valid, and with increasing the magnetic field the gap is eventually suppressed as $B^{-1}$. It thus seems that a favorable condition for the realization of the Majorana-Weyl transition is present. As often forgotten, Majorana zero-energy modes are, by themselves, massless excitations (in a fully relativistic approach, the energy cannot be zero for a non-zero mass), but an effective mass is generated dynamically by  scalar-fermion interactions \cite{Rossi}. While discussing heterostructures with possible Majorana modes, the realization of the condition in (\ref{TS}) is, thus, only one basic requirement, the other one being a non-vanishing $\Delta$: with a vanishing induced gap, no Majorana fermion emerges, even if the condition (\ref{TS}) is fulfilled.

If the maximal superposition takes place, the system should exhibit parity violation, and this prediction could be checked experimentally (for example by measuring the $I$-$V$ characteristic; see \cite{Nobukane}). However, such assumption can be easily (and more realistically) relaxed. For higher values of the magnetic field, the coherence length increases in any case, this producing an effective overlap between the Majorana modes localized at the two ends of the wire, so that the excitations in the systems can be described by a state function of the form $\alpha \psi_{M1} + \beta \psi_{M2}$, with two (in general) non-equal $\alpha. \beta$ coefficients: in such a case, we have not one pure right-handed or left-handed state, but rather a suitable mixture of them. The point, however, remains: the system is described in terms of (massless) Weyl fermions rather than (massive) Majorana fermions, since the gap is suppressed under given conditions (higher values of the magnetic field).

Now, is it possible to further check experimentally such a scenario or, more in general, to distinguish Majorana from Weyl fermions?

As it is well-known, as a consequence of  their basic property of self-charge conjugation, Majorana fermions do not support $U(1)$ symmetry of  any kind, so that the corresponding charge is identically zero \cite{KayserNieves}. Electromagnetic interaction of Majorana fermions is, then, severely constrained. As a consequence of Lorentz invariance, the most general expression for the conserved relativistic electromagnetic current for a spin-1/2 fermion may be written (in the momentum space) as \cite{KayserNieves}:
\bea 
& & \!\!\!\!\!\!
\ov{\psi}_M(k') J_\alpha(q) \psi_M(k) =  \nonumber \\ & & ~~~ ~~~ 
\ov{\psi}_M(k') \left[ e \gamma_\alpha - i \, \frac{\mu}{2 m} \, \sigma_{\alpha \beta} q^\beta - i \, \frac{d}{m} \, \sigma_{\alpha \beta} q^\beta \gamma_5 
\right. \nonumber \\ & & \left. ~~~ ~~~ 
+ \frac{a}{m^2} \left( q_\alpha q_\beta - g_{\alpha \beta} q^2 \right) \gamma^\beta \gamma_5 \right] \psi_M(k) , 
\label{nuemc}
\eea
where $q=k'-k$, $g_{\alpha \beta}$ is the metric tensor with signature $(+,-,-,-)$, $\sigma_{\alpha \beta} = (i/2) [\gamma_\alpha, \gamma_\beta]$ and $\gamma_\alpha$ are the usual set of Dirac gamma matrices. The first term in square brackets represents the electric charge contribution $\ov{\psi}_M \gamma_0 \gamma_\alpha \psi_M = (\gamma_0 \gamma_\alpha)_{ab}\psi_{Ma} \psi_{Mb}$. Since the matrix $\gamma_0 \gamma_\alpha$ is symmetric under the exchange of the spinor indices $a$ and $b$, while the fermion bilinear $\psi_{Ma} \psi_{Mb}$ is antisymmetric, such contribution identically vanishes, as expected. The same is true for the magnetic (second term) and electric (third term) dipole moment contributions, so that Majorana fermions cannot couple at all to a real photon field. Nevertheless, the field $\psi_M$ may couple to a {\it virtual} photon field through an anapole moment \cite{Zeldo} (fourth term) which, in the non-relativistic limit, reduces to
\bea
& & \!\!\!\!\!\!\!\!\!
\frac{a}{m^2} \left( q_\alpha q_\beta - g_{\alpha \beta} q^2 \right) \gamma^\beta \gamma_5 \rightarrow 
\nonumber \\ & & ~~~ ~~~ 
\frac{a}{m^2} {\mathbf q}^2 \left[ \sigmabf - \frac{{\mathbf q}}{|{\mathbf q}|} \left( \frac{{\mathbf q}}{|{\mathbf q}|} \cdot \sigmabf \right) \right] = \frac{a}{m^2} {\mathbf q}^2 \sigmabf_{\perp}.  
\label{nonrel}
\eea
{Since the electromagnetic current in (\ref{nuemc}) has to be finite for $m$ tending to zero, it follows that the magnetic $\mu$ and electric $d$ dipole moments, as well as the anapole moment $a$,} vanish in the limit of zero mass, so that a (neutral) Majorana fermion differs from a Weyl fermion just for the anapole moment interaction. Such a property can, then, be used to distinguish between the two cases.

In condensed matter systems, in addition to the usual magnetic and electric dipole, the anapole moment emerges as a different type of vector order parameter, which describes the distribution of spins and electric currents in the system. {Note that, from (\ref{nonrel}), the current is transverse and spin-dependent: the spin vector $\sigmabf$ of the fermion 
forms a spin spiral leading to a toroidal moment.} Indeed, in classical electromagnetism, a non-zero anapole moment can be generated by a toroidal current winding \cite{Haxton}: the current in a solenoid folded into a torus induces a circular magnetic field inside the solenoid, which leads to an anapole moment perpendicular to the magnetic field. As it is evident also from this example, in general the anapole moment changes sign under both space inversion and time reversal, so that it requires either the introduction of an axial vector component into the current or a parity admixture in the ground state of the given system. Since electromagnetic interactions are invariant under both space and time inversion, in solid state systems the anapole moment should appear as the result of a spontaneous breaking of these symmetries. This can be obtained, for example, in systems exhibiting classical spin ordering, as well as in frustrated quantum spin systems, whose ground state degeneracy leads to states with spontaneously broken space and time symmteries, or, also, by means of spin vortices \cite{Spaldin}. 

Another relevant property is that, due to the $q_\alpha q_\beta - g_{\alpha \beta} q^2$ structure of the anapole term in the current (\ref{nuemc}) and to the current conservation $q_\alpha J^\alpha =0$, the anapole moment does not interact with the external electromagnetic field $F_{\alpha \beta}$, but only with its derivative $\partial^\beta F_{\alpha \beta}$ or, rather, with the external current $j_{\rm ext}$. Indeed, the interaction potential due to the anapole moment may be written as $J_\alpha A^\alpha = S^\alpha \partial^\beta F_{\alpha \beta}$, where $S^\alpha = (a/m^2) \gamma^\alpha \gamma_5$, or, by employing the Maxwell equations, $J_\alpha A^\alpha = S_\alpha j_{\rm ext}^\alpha$. This gives rise to a contact interaction, which allows effective electromagnetic interaction with electrons in the system only to the extent that their wavefunctions penetrates  the Majorana fermion \cite{Haxton} (in the classical picture, the interaction with a test particle can only take place if the particle penetrates the torus).

\begin{figure}
\begin{center}
\epsfig{height=3.5truecm,file=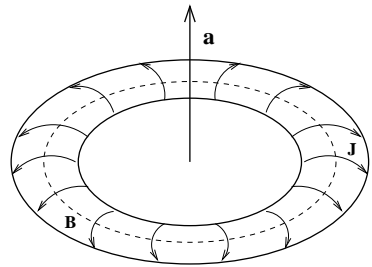} %
\caption{{Topologically invariant edge currents, induced on a torus by a circular magnetic field along the internal loop of it, generate an anapole moment.}}
\label{fig1}
\end{center}
\end{figure}

{A simple toy model showing the insurgence of a magnetic anapole interaction is that of an electron moving on a ring (centered at the origin of an $xy$ plane) in a circular magnetic field of the form ${\bf B} =(-B \sin \phi, B \cos \phi, 0)$, where $\phi$ is the polar angle. The electromagnetic interaction for such a system is described by an hamiltonian term proportional to the cross product ${\bf a} \sim {\bf r} \times {\bf s}$ \cite{Zelevinsky}, $\bf s$ being the spin vector, which evidently changes the parity of the given state of the system and can only exist in quantum states with non-zero spin. Similar considerations apply to Majorana fermions, which are essentially Bogoliubov quasiparticles with equal superpositions of electrons and holes, so that their anapole moment can be induced just as an electromagnetic moment corresponding to an effective toroidal current. Indeed, topologically invariant edge currents may be induced on a torus by a suitable magnetic field (see Fig. \ref{fig1}): if the torus is just placed in an homogeneous field along the torus axis, it would not generate a permanent circulation of the appropriate type, while an anapole moment is induced by a (necessarily inhomogeneous) magnetic field that would run along the  internal loop of the torus. Alternatively, however, an homogeneous magnetic field along the torus axis would suffice if the toroid is chiral, since the current which goes around the central hole of a toroidal chiral tube follows a spiral pattern with the required topological properties \cite{Lijnen}.}

\begin{figure}
\begin{center}
\epsfig{height=6truecm,file=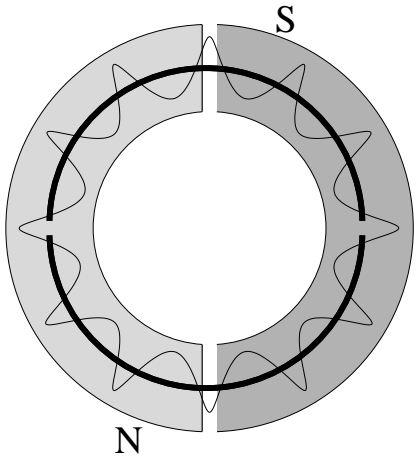} %
\caption{A possible geometry for detecting Majorana fermions (and distinguish them from Weyl fermions) by means of their magnetic anapole moment: two semi-circular nanowires are placed over a normal (N) $-$ superconductor (S) junction with the same symmetry, and a current wound around the nanowires produces a circular magnetic field.}
\label{fig2}
\end{center}
\end{figure}

In the experimental setups considered above, where zero-bias conductance peaks have been observed, it is likely not possible to test our interpretation in terms of a transition from Majorana to Weyl states, since the topology employed does not allow a non-vanishing anapole moment, so that no effect manifests both in the Weyl phase and in the Majorana one. In general, a vortex spin structure is required in solid state systems, since spin vortices like planar quadratic or hexagonal head-to-tail arrangements of spins carry an anapole moment \cite{ChupisRev}. Here, the first key ingredient is to have not a rectilinear wire, but rather a toroidal one (for an example of possible geometries, in the realm of molecules, see \cite{Ceulemans}): it may be composed of two semi-circular nanowires \footnote{In principle, as noted in \cite{Kitaev}, an open string is not strictly required in order to have Majorana fermions, a closed loop being suitable as well.} placed, with the corresponding edges opposite one another, over the normal/superconducting electrodes, as shown in Fig. \ref{fig2}. A circular magnetic field that follows the loop of the torus is, then, required; in the presence of a system with chiral symmetry, even a uniform magnetic field along the principal axis of revolutions may fit well \cite{Ceulemans}. 

Although toroidal interactions are much weaker than those of dipole type, in such a complex heterostructure it would in principle be possible to detect effects produced (or not) by the anapole moment of the Majorana (Weyl) fermions, these effects being directly related to the energy shift induced by the interaction with the anapole moment \cite{Lewis}. If the electron current interacting with the anapole moment of the Majorana fermions is a spin current, the effective interaction potential can be cast in the form \cite{Flam}:
\be  \label{anapol}
H_{\rm eff} = 2 k_A \, {\mathbf n} \times {\bf s} \cdot {\bf S} ,
\ee
where $\bf s$, $\bf S$ are the spins of the electrons and Majorana fermions, respectively, $\bf n$ is the unit vector directed along the axis of the spin vortex, and $k_A$ is a coupling constant. In a sense, it is similar to a spin-orbit interaction, where the role of the ``orbit'' is played by the vortex, and with the important difference that, here, there is no contribution from the components of $\bf s$ or $\bf S$ parallel to the vortex axis. The simple Jackiw-Rossi \cite{Rossi} or Fu-Kane \cite{FuKane} 2D system is now modified by the presence of the additional term in (\ref{anapol}); by choosing, for example, the vortex axis along the $z$-axis and the electron spin current in the $x$ direction, the hamiltonian \cite{Nishida} describing the system is given by
\be \label{hana}
H = \left[ \ba{cc} \sigmabf \cdot {\mathbf p} +V - \mu  & \Delta \\ \Delta^\ast & - \sigmabf \cdot {\mathbf p} + V + \mu \ea \right] ,
\ee
where $V=\sigma_z E_Z + \sigma_y k_A$. The most intriguing consequence of such modification is that the 
condition in (\ref{TS}) is now replaced by the following one:
\be \label{TSm}
E_Z > \sqrt{\Delta^2 + \mu^2 - k_A^2}\,  . 
\ee
The presence of the Majorana anapole contribution thus results into a {\it lowering} of the critical magnetic field required to have a zero-bias conductance peak. 

The scenario is now clear. By performing experiments similar to that described in \cite{Mourik} (and \cite{Das}), but with the geometry depicted in Fig. \ref{fig2}, the appearance of a zero-bias conductance peak is registered for a given value $B_{c1}$ of the applied critical field along the wire, when the circular magnetic field (within the torus) is switched off. However, provided that such circular field is then switched on, thus inducing an anapole moment to the Majorana fermions, a zero-bias peak is now registered for a value $B_{c2}$ of the critical field that is {\it lower} than the previously measured $B_{c1}$. This is a very distinctive feature of the presence of Majorana fermions and, moreover, of their anapole interaction. Of course, this does not apply to Weyl fermions \footnote{The limiting case $k_A=\Delta =0$ (Weyl fermions) is phenomenologically different from the case with  $k_A=\Delta \neq 0$ (Majorana fermions), since the superconducting properties of the structure are different. It is, however, intriguing the fact that a sufficiently high anapole moment (if effectively realizable in the systems considered here) may mimick Majorana fermions in ``absence'' of superconductivity.}.

Summing up, we have given a thorough interpretation of the apparent splitting of the zero-bias conductance peak, seen in recent experiments aimed at revealing the presence of Majorana fermions in complex heterostructures. We have argued that such splitting, observed for higher values of the applied magnetic field along the nanowire, marks the transition from (massive) Majorana fermions to (massless) Weyl fermions, which can also be viewed (according to common thinking) as a superposition of the two Majorana fermions present at the opposite edges of the nanowire. In order to discriminate among this interpretation and that in terms of a coupling of two nearby Majorana modes (no Weyl fermions), we have proposed to search for distinctive properties of Majorana fermions with respect to Weyl ones. They are related to the only possible electromagnetic interaction allowed to massive Majorana particles, i.e. induced by a non-vanishing anapole moment. We have, then, showed that in suitably designed heterostructures, similar to those employed in existing experiments but with a toroidal symmetry, the additional anapole interaction manifests itself in the lowering of the critical magnetic field required for the appearance of a zero-bias conductance peak, this phenomenon being ruled by external control parameters. 

It is as well remarkable that anapole interaction of Majorana fermions, not previously considered, is also a powerful method to control their dynamics and, thus, potentially relevant in applications for manipulating 
the qubit state in quantum computation. Further studies in this novel direction will certainly throw new light on this exciting subject.


\end{document}